# Impacts of annealing on structural and photophysical properties of Zinc Phthalocyanine adsorbed on graphene[1]

Gautier Creuzer[1,2], Quentin Fernez[3], Nataliya Kalashnyk[4], Zohreh Safarzadeh[5], Lydia Sosa Vargas[3], Céline Fiorini-Debuisschert[1], Nicolas Fabre*[1] and Fabrice Charra[1]

Address: [1]Université Paris-Saclay, CEA-CNRS, Service de Physique de l'État condensé (SPEC), F-91191, Gif-sur-Yvette, France, [2]Present address: Laboratoire Kastler Brossel, Collège de France, CNRS, ENS-Université PSL, Sorbonne Université, 11 place Marcelin Berthelot, F-75231 Paris, France

[3]Sorbonne Université, CNRS, Institut Parisien de Chimie Moléculaire (IPCM), F-75005, Paris, France, [4]Univ. Lille, CNRS, Centrale Lille, Univ. Polytechnique Hauts-de-France, UMR 8520-IEMN – Institut d'Electronique, de Microélectronique et de Nanotechnologie, F-59000 Lille, France and [5]Sorbonne Université, CNRS, Laboratoire Physico-Chimie des Électrolytes et Nano-Systèmes Interfaciaux (PHENIX), F-75005, Paris, France

* Corresponding author

Email: Nicolas Fabre – nicolas.fabre@cea.fr

# Abstract

We report the demonstration and analysis by combined Scanning-Tunneling-Microscopy (STM) and optical microspectroscopy of a 2D phase change experienced by a self-assembled zinc phthalocyanine (ZnPc) monolayer adsorbed on graphene. To probe the intrinsic properties of individual ZnPc molecules, they are spatially confined within the pores of a self-assembled 2D matrix. This confinement allows us to track a phase change induced by annealing, that we discuss in terms of a planar square to shuttlecock molecular transition. We show that after annealing of adsorbed ZnPc, the exposition of Zn atoms to reactants in a supernatant solution is improved *e.g.* for metal-ligand formation towards 3D self-assembly.

# Keywords

# Introduction

Combining the properties of graphene and molecular semiconductors in a given material organized at the molecular scale appears a promising route to design original and innovative electronic devices [1] such as diodes [2,3], transistors [4-6], photodetectors [7,8], solar cells [9-11], or light-emitting devices [12]. In such heterostructures most electronic processes take place at the interface between graphene and molecular media, and are strongly influenced by various structural parameters at the molecular scale [13]. For example, face-on or edge-on orientation of π-conjugated molecules on graphene turns on or off π-stacking interactions with large consequences on hole conductivity and rectifying properties [14]. It is thus important to develop a deep understanding about how to control the organization of the π-conjugated molecules that are in contact with graphene and how this organization impacts their electronic excitations.

In this context, porphyrins, phthalocyanines (Pcs) and their metalated complexes (MPcs), a well-known family of organic semiconductors, have been the subject of intensive research [15]. This family of molecules offers many advantages for industrial applications such as nontoxicity, thermal and chemical stability, and strong optical absorption [16,17]. It has also become a paradigm in fundamental research on organic semiconductors. The flexibility offered by the choice of the coordinated central metal cation permits to vary their electronic, photonic and spin-related properties. It also influences their stacking geometry, in particular as a result of the change in relative stability of planar square *versus* shuttlecock shapes depending on the central atoms [18].

As bulk materials, phthalocyanines have long been known to exhibit several polymorphs [19] with marked spectral differences in the Q-band range [600 nm–800 nm] [20]. The transition between polymorphs can be controlled by thermal treatments, as often shown for example with α and β phases of zinc phthalocyanines (ZnPc), with consequences on molecule orientations relative to the substrate and absorbance efficiency [21] or dynamics of charge migration and charge transfer to substrate [22]. Even inside a given phase, minute structural variations in phthalocyanine-based materials can strongly impact their absorption and luminescence spectra, in particular by allowing or not the formation of intermolecular Frenkel charge-transfer excitons [23].

2D assemblies of self-organized adsorbed conjugated molecules on graphene has attracted particular interest since they permit to focus on the specific properties of the molecules that are in direct contact with graphene. These studies benefit from in-depth analysis offered by scanning probe techniques combined with measurements of optical absorption [24,25], photoluminescence [26], or graphene-enhanced Raman scattering [27]. Such combinations allow atomic-scale inspection of both molecular organization and electronic structures. These 2D systems are accessible to numerical simulations such as Density-Functional Theories, in particular as concerns Pcs [18,28]. Such simulations can be compared with results of STM and Scanning Tunneling Spectroscopy (STS) [29]. Similarly to their thicker 3D counterparts, 2D metal-free Pc [30] or metalated Pc [31] assemblies adsorbed on various substrates have shown important phase changes induced by thermal treatments. In particular, ZnPc has been shown to present such 2D phase transitions on various substrates such as TiO2 [32], Au(111) [33,34], InSb [35], or ZnS [36]. These studies have emphasized the diversity of mechanisms involved in the relative stability of 2D phases, including intermolecular versus molecule-substrate forces, face-on versus edge-on molecule orientation, planar square versus shuttlecock molecule shape, *etc.* Although the importance of thermal treatment on the interfacial structure of $H_2$Pc [37] and ZnPc [38] on highly oriented pyrolytic graphite (HOPG) and graphene has been demonstrated, its mechanisms, application in interface structure management, and impact on electronic or photonic properties are still poorly understood.

In this paper, focusing on ZnPc, we report the demonstration and analysis by combined STM operated at the air-solid interface and optical microspectroscopy of a 2D phase change experienced by ZnPc self-assembled monolayer on graphene or HOPG. Single ZnPc molecules are guest-isolated within the nanocavities of a self-assembled 2D host matrix, preventing intermolecular interactions and allowing for individual characterization. This phase change, induced by annealing, is discussed in terms of a planar square to shuttlecock molecular transition and its beneficial consequences on 3D metal-ligand formation are shown.

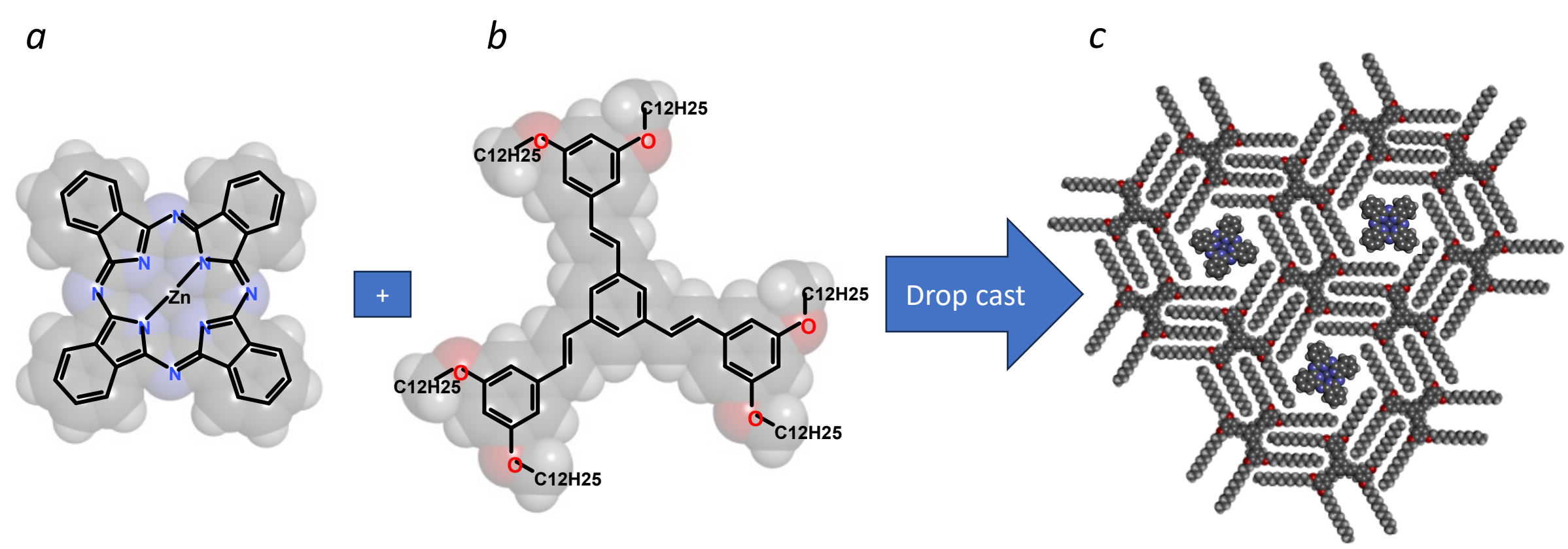

**Scheme 1:** Molecular structures of the TSB35-C12 host-matrix molecule (*a*) and of the Zinc Phthalocyanine (ZnPc) guest molecule (*b*), together with the self-assembled pattern obtained after depositing successively the host and guest species (*c*).

# Results and Discussion

The samples studied consisted in ZnPc molecules (scheme 1*a*) individually embedded into the pores of a 2D self-assembled network of 1,3,5 tristyrylbenzene substituted by dodecyl alkoxy peripheral chains TSB35-C12 [39] (scheme 1*b*) grown by drop-casting from a toluene solution either on HOPG, for STM experiments, or on a transparent monolayer Chemical-Vapor-Deposition (CVD) graphene on glass substrates, for optical measurements. This system has been shown to form highly reproducibly honeycomb guest-host monolayers (scheme 1*c*, denoted ZnPc:TSB35-C12) that are robust at air under ambient conditions [40] and in which the guest ZnPc molecules are confined at a distance of 4.3 nm center-to-center from each other [39]. The measurements were performed on the dried samples both before and after annealing at 80°C for 30 min or 180 min.

## Optical absorption micro-spectroscopy

Optical absorption was measured by transmission micro-spectroscopy for ZnPc:TSB35-C12 grown on a highly transparent CVD graphene monolayer transferred on a microscope glass cover plate. The results are reported in Figure 1. Before annealing (Figure 1*a*, blue curve), we observe the ZnPc Q-band peak at 676 nm, close to the peak absorption in solution (671.5 nm in toluene [41]). During the 3-h annealing at 80°C, this Q-band peak turns into a new band peaking at 711 nm (Figure 1*a*, red curve). This is accompanied by a parallel spectral displacement of its vibronic replica, located 0.21 eV higher in energy, from 608 nm to 630 nm. At intermediate time of annealing (30 min, Figure 1*a*, purple curve) the two distinct Q-band peaks at 676 nm and 711 nm are observed simultaneously, showing that the annealing produces a replacement of one peak by the other rather than a continuous spectral shift of the peak. One can notice a small shoulder at 711 nm already present before any annealing (Figure 1*a*, blue curve). The hyperspectral 1D image (Figure 1*b*) shows that at optical spatial resolution (~1 µm) both Q-band absorptions are uniformly distributed all along the sample considered area. This holds also for the intermediate situation, showing that the spectral effect of annealing does not involve a 2D phase transformation with domains larger than the micrometer.

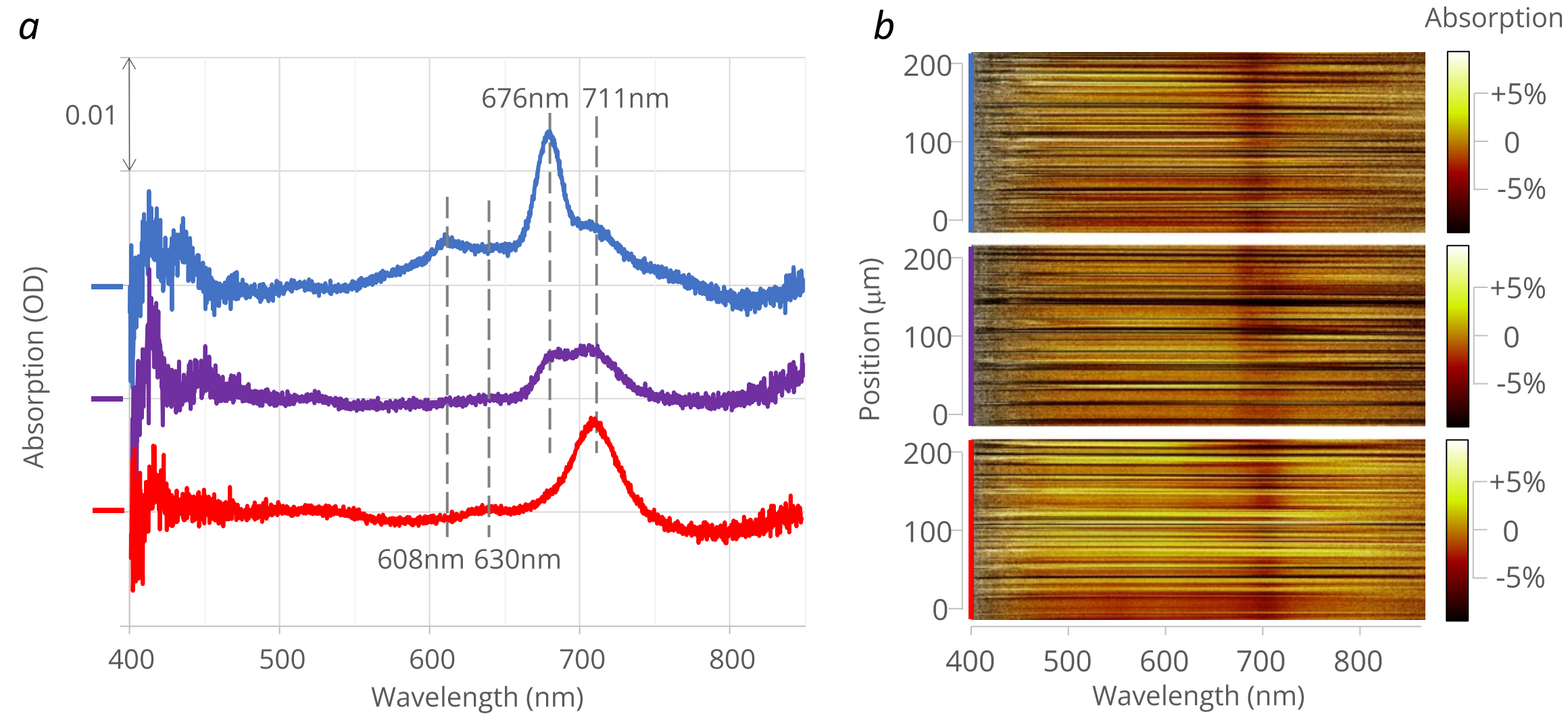

**Figure 1:** *a*: micro-absorption spectra of a ZnPc:TSB35-C12 self-assembled monolayer on CVD graphene before (blue), and after 30-min annealing (purple) and 3-h annealing (red) at 80°C. Baselines corresponding to the average graphene absorption has been subtracted and the curves have been vertically shifted for clarity. *b*: spatial variations of the corresponding transmission spectra, represented in pseudocolor, as recorded along a 230-µm line on the sample. From top to bottom: before annealing (blue), after intermediate annealing (purple), after complete annealing (red). The Q-band absorption peaks before and after annealing (676 nm and 711 nm, respectively) are indicated on the spectra, together with their vibronic replicas (608 nm and 630 nm, respectively)

## Fluorescence and Raman micro-spectroscopy

The strong excited-state quenching exerted by the graphene layer on adsorbed ZnPc molecules permits to strongly reduce the fluorescence quantum yield of the latter and thus permits the measurements of its Raman scattering spectrum. The corresponding spectra excited at 633 nm are reported in Figure 2. Only ZnPc and graphene absorb at this excitation wavelength and can emit a resonantly-enhanced Raman scattering detectable even at the monolayer level. On the contrary, TSB35-C12 is driven far from its resonance (its absorption starts at wavelengths shorter than 350 nm) and it doesn't emit any detectable Raman signal. Moreover, Raman signal of adsorbed phthalocyanines benefits from large Graphene-Enhanced Raman Scattering (GERS) [42].

The signal amplitude corresponds to the sum of fluorescence and Raman scattering, the dominant inelastic scattering sources. A residual fluorescence is still observed as a large background covering the [600-2500] $cm^{-1}$ region, *i.e.* [660-750] nm in terms of emitted wavelength, with a quantum yield less than $10^{-6}$ as roughly estimated by comparison with the typical signal obtained from a reference standard composed of a rhodamine 6G doped PMMA thin film. This fluorescent background appears attenuated after annealing, which corresponds to an improved quenching by the graphene and suggests an increased interaction of the ZnPc $\pi$-conjugated electrons with graphene.
As concerns the Raman scattering, the pattern formed by the three main peaks in the range 1300 – 1600 $cm^{-1}$ is characteristic of ZnPc response [43]. However, the highest energy peak is strongly shifted from its value of 1506 $cm^{-1}$ in the bulk to 1543 $cm^{-1}$ in this assembly. The same shift has been reported for the ZnPc embedded inside carbon nanotubes or adsorbed on their surface [44]. This is considered as a signature of $\pi$-stacking of the Pc conjugated core on the nanotube surface and its observation here is consistent with an adsorption of ZnPc on graphene, the molecule lying flat on the surface.
The frequencies of these three peaks are similar before and after annealing (1543 $cm^{-1}$, 1472 $cm^{-1}$ and 1374 $cm^{-1}$), within our experimental accuracy (±5 $cm^{-1}$) limited by the widths of these peaks. However, the two highest-frequency peak intensities are markedly reduced by about 50%, as roughly estimated accounting for fluorescence baseline, whereas the intensity of the lowest-frequency peak remains unchanged.

The vibration at 1543 $cm^{-1}$ ($B_{1G}$ symmetry) and that at 1472 $cm^{-1}$ ($A_{1G}$ symmetry) both involve out-of-plane stretching of C-N-C bonds of the Pc conjugated ring and out-of-plane bending of peripheral C-H [43]. In contrast, the vibration at 1374 $cm^{-1}$ ($B_{1G}$ symmetry) corresponds to in-plane stretching of C-C bonds of the Pc conjugated ring and in-plane bending of C-H. Such intensity changes in the bands corresponding to out-of-plane vibrations have been reported to be a consequence of an axial displacement of the zinc ion and subsequent out-of-plane distortion of the ring [43]. It results from a lowering of the symmetry of ZnPc from D4h in the planar square geometry to C4v in the shuttlecock shape.

We can thus conclude that our observed changes in fluorescence and Raman scattering are consistent with a change from an initial planar square geometry already present in majority in solution to a shuttlecock geometry, with the central Zn atom pointing outward, *i.e.* away from graphene, thus allowing the Pc conjugated structure to approach closer to the graphene substrate.

Returning to the absorption measurements, we infer that this planar square to shuttlecock transition is accompanied by a shift of the Q-band, a small proportion of molecules in solution being already under shuttlecock form.

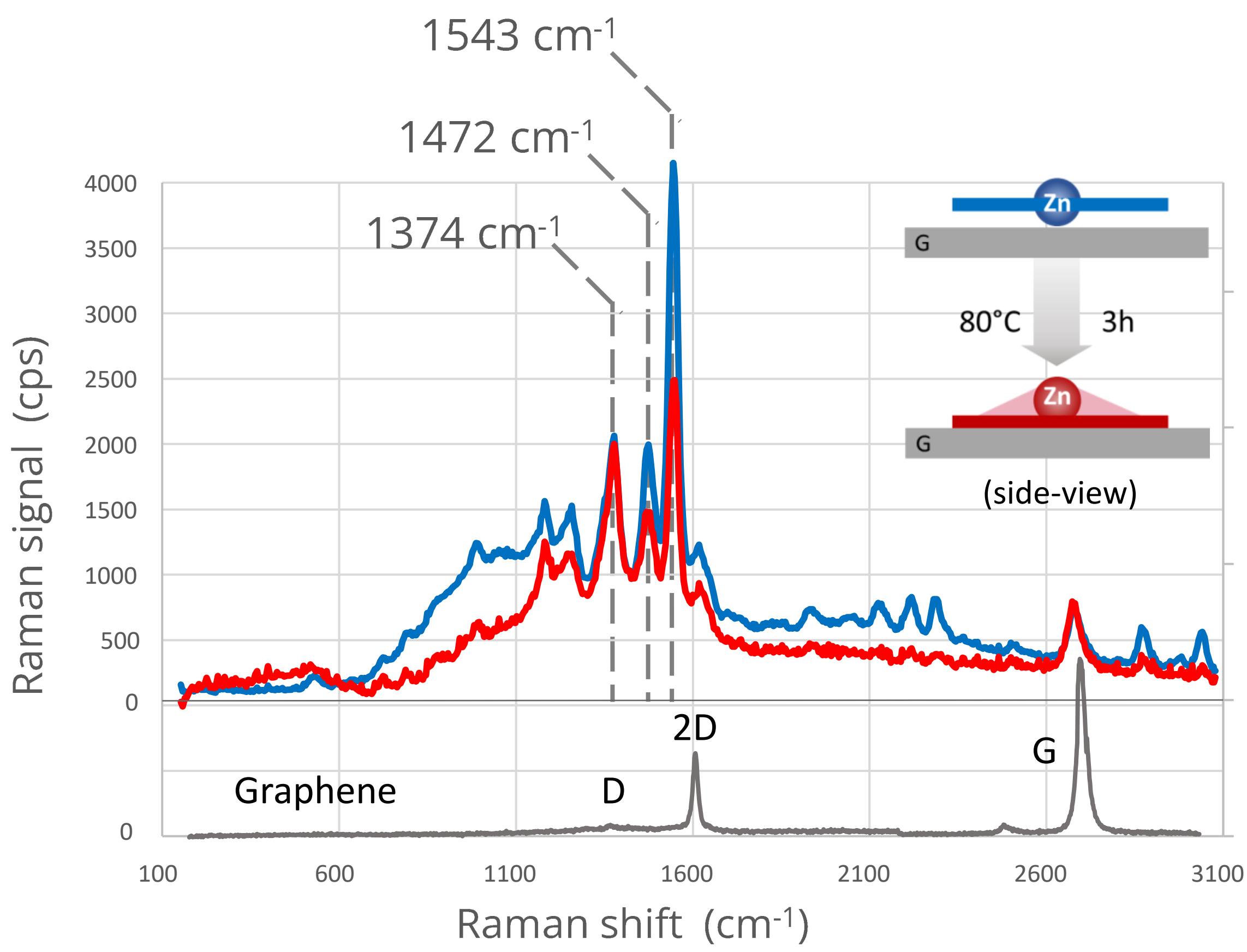

**Figure 2:** Raman scattering microspectroscopy excited at 633 nm for a ZnPc:TSB35-C12 self-assembled monolayer on CVD graphene before (blue) and after (red) 3-h annealing at 80°C. The Raman photon counts are acquired with the same acquisition time and excitation intensity so that the amplitudes can be compared. The main three central peaks (1543 $cm^{-1}$, 1472 $cm^{-1}$ and 1374 $cm^{-1}$) discussed in the text are highlighted. Neat graphene response before ZnPc and TSB35-C12 deposition (grey), for reference. The standard D, 2D and G peaks are labelled. The insert illustrates the proposed planar-square to shuttlecock transition (G: graphene substrate).

## Scanning Tunneling Microscopy

Although the TSB35-family molecules form the same atomically-precise self-assembly patterns on graphene and HOPG, the former is less convenient for STM observations because of defects inherently present in the transferred CVD graphene [45]. We thus studied the effects of annealing on the structure of ZnPc trapped in a TSB35-C12 nanoporous matrix by STM on HOPG, as reported in Figure 3, the same behavior being expected on graphene, in line with our previously published results [40,45].

The annealing brings mainly two changes. Firstly, as observed by comparing Figure 3*a* and 3*d*, the average size of the domains increases and disordered areas tend to disappear. Such trends are expected for self-assembled monolayers in general and have been reported and analyzed in detail in the specific case of the TSB35 monolayers [46]. Given the robustness of the TSB35-C12 matrices, it is not surprising that such behaviors are preserved in the presence of trapped ZnPc. The lattice constant, 4.3 nm, is also preserved, as imposed by the TSB35-C12 matrix [39]. Secondly, and more pointedly, the high-resolution STM images show a marked evolution internally to the network unit cell, upon annealing (compare Figure 3*b* and 3*e*). Mainly, the brighter spots, attributable to ZnPc molecules, become smaller and, consequently, the less bright TSB35-C12 conjugated cores become more apparent around the ZnPc spots (see insert in Figure 3*d* for spot ascription). This is further substantiated by comparing the profiles in Figures 3*c* and 3*f*, the latter showing thinner bumps above ZnPc and more visible bumps above matrix molecules. It is noticeable that before annealing (Figure 3*b*) a few trapped ZnPc already appear smaller, with a pattern similar to the majority ones in Figure 3*e*. This change in ZnPc imaging is consistent with the interpretation in terms of planar square to shuttlecock structure mentioned before. Actually, the Pc conjugated core being brought-in deeper into the matrix, *i.e.* closer to the substrate, the ZnPc STM image recenters on the central Zn atom and the visibility of the TSB35-C12 conjugated cores, relative to ZnPc, is improved. Such a better resolution, that is a smaller spot size produced by the central

Zn atom, is also characteristic of reduced fluctuations through a tighter anchoring of the phthalocyanine on HOPG inside the TSB35-C12 matrix.

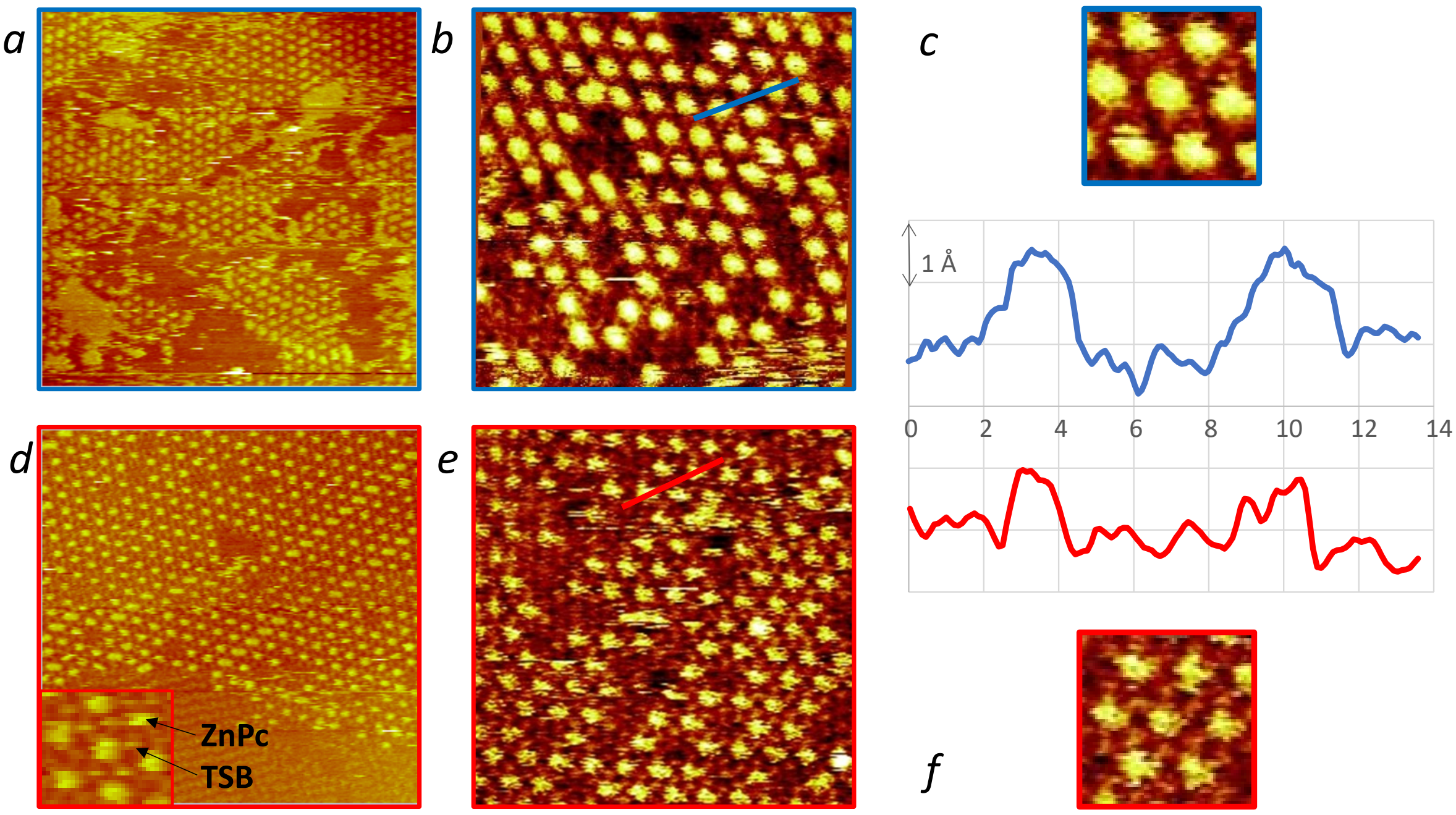

**Figure 3:** Scanning Tunneling Microscopy (STM) images showing the molecular network of a self-assembled monolayer of ZnPc:TSB35-C12 on HOPG, as measured before (*a*, 130×130 nm²; *b*, 50×50 nm², blue-framed) and after (*d*, 90×90 nm²; *e*, 50×50 nm², red-framed) annealing (3 h at 80°C.). An insert in *c* shows the STM spots ascription on a zoomed detail. Selected STM profiles along lattice axes crossing both ZnPc and TSB molecules are also represented (*c*; *f*), together with enlarged details of high-resolution images *b* and *e*. The current setpoint was ~3 pA and the sample bias was -0.9 V. The images are not corrected for the thermal drift.

## Absorption anisotropy

Finally, incidence-angle dependence of the optical absorption in transverse magnetic polarization have been measured before and after annealing, as reported in Figure 4.

As illustrated in the figure, by varying the incidence angle $\theta$ in *p* polarization, the optical electric field varies from in-plane (normal incidence, $\theta$ =0) to out of plane (grazing incidence, $\theta \rightarrow 90°$). The absorption being maximum when molecular transition dipole moments of the probed optical transition (here the Q-band) are parallel to the electric field, this experiment permits an evaluation of molecular orientation [25,47,48]. This analysis clearly shows that the vectors of transition dipole moments for the Q-band peak of ZnPc are oriented in-plane both before and after annealing. With respect to molecule geometry, the transition dipole moments for the Q-band transition of phthalocyanineis oriented in the molecule $\pi$-conjugated plane ($Q_x$ and $Q_y$ degenerate transitions [49]). The presently observed anisotropy of the Q-band absorption thus shows that ZnPc molecules are oriented face-on relative to the substrate, both before and after annealing. This is consistent with our interpretation of ZnPc molecular transition while the molecule remains adsorbed on graphene or HOPG, within TSB pores.

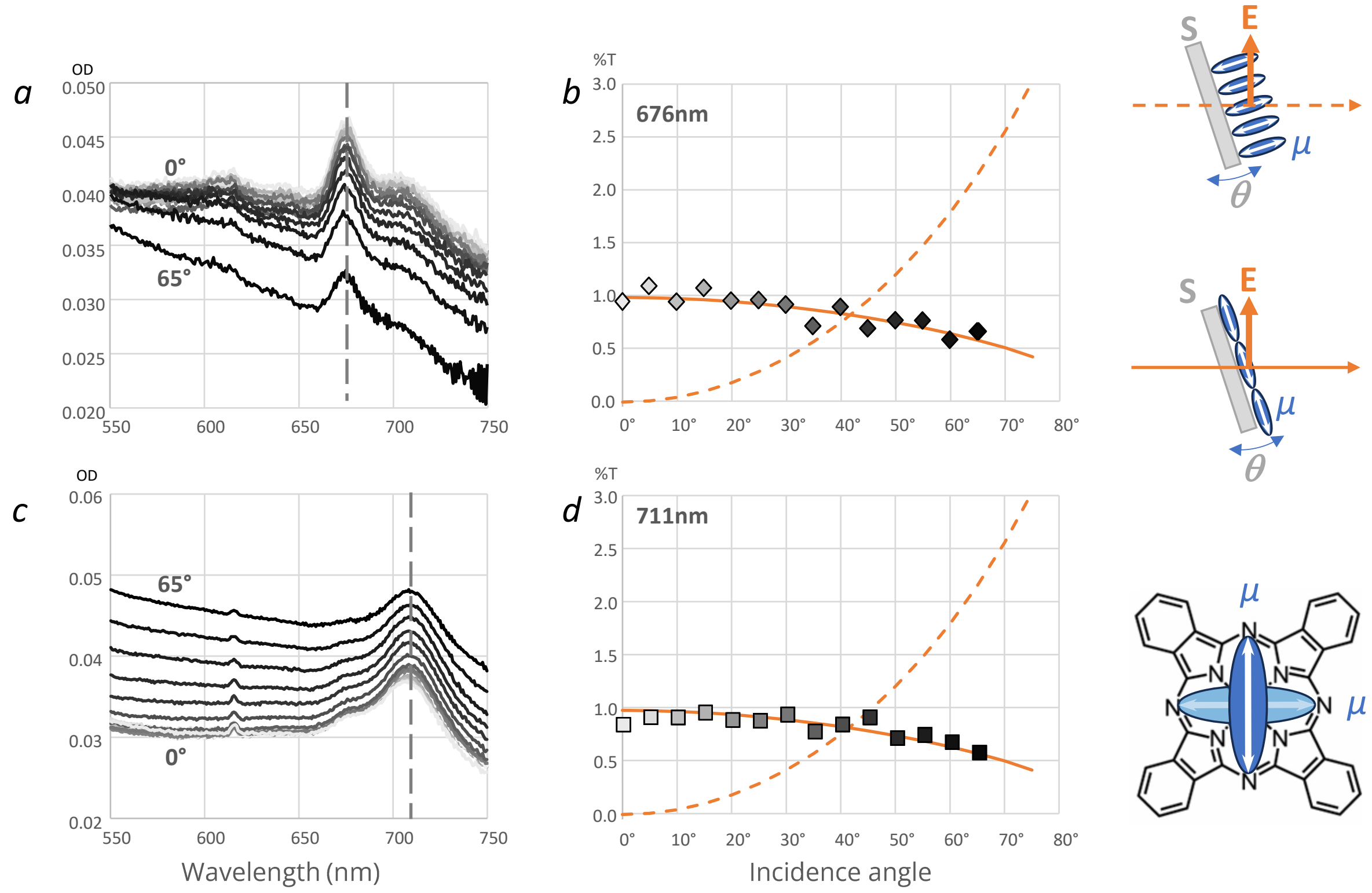

**Figure 4:** Incidence-angle dependence, as measured every 5° from 0° (normal incidence, light grey) to 65° (black), of *p*-polarized (TM) absorption of a self-assembled monolayer of ZnPc:TSB35-C12 on CVD graphene before (*a*) and after (*c*) annealing (3 h at 80°C) ; corresponding variations of ZnPc Q-band peak amplitude at maximum before (*b*, 676 nm) and after (*d*, 711 nm) annealing. Solid and dotted lines in (*b*, *d*) represent theoretical relative variations of angle-dependent absorption for, respectively, a transition dipole moment oriented either in-plane or normal to the substrate plane. The corresponding orientation hypothesis and measurement geometry are also schematized. Horizontal arrows: beam propagation direction, vertical arrow: optical electric field, S: substrate plane tilted by the variable angle $\theta$. Blue ellipsis ($\mu$): ZnPc transition dipole moment.

## Metal-ligand complex formation

The above hypothesised planar-square to shuttlecock transition is expected to alter the exposition of the central Zn atom of adsorbed ZnPc to reactants in a supernatant solution. This can be expected to influence the formation of metal-ligand coordination complex of an adsorbed ZnPc with pyridyl groups. Such a process has been reported recently using pyridyl-functionalized perylenes (BPDI4Py, Figure 5, see details in the cited reference) [50].

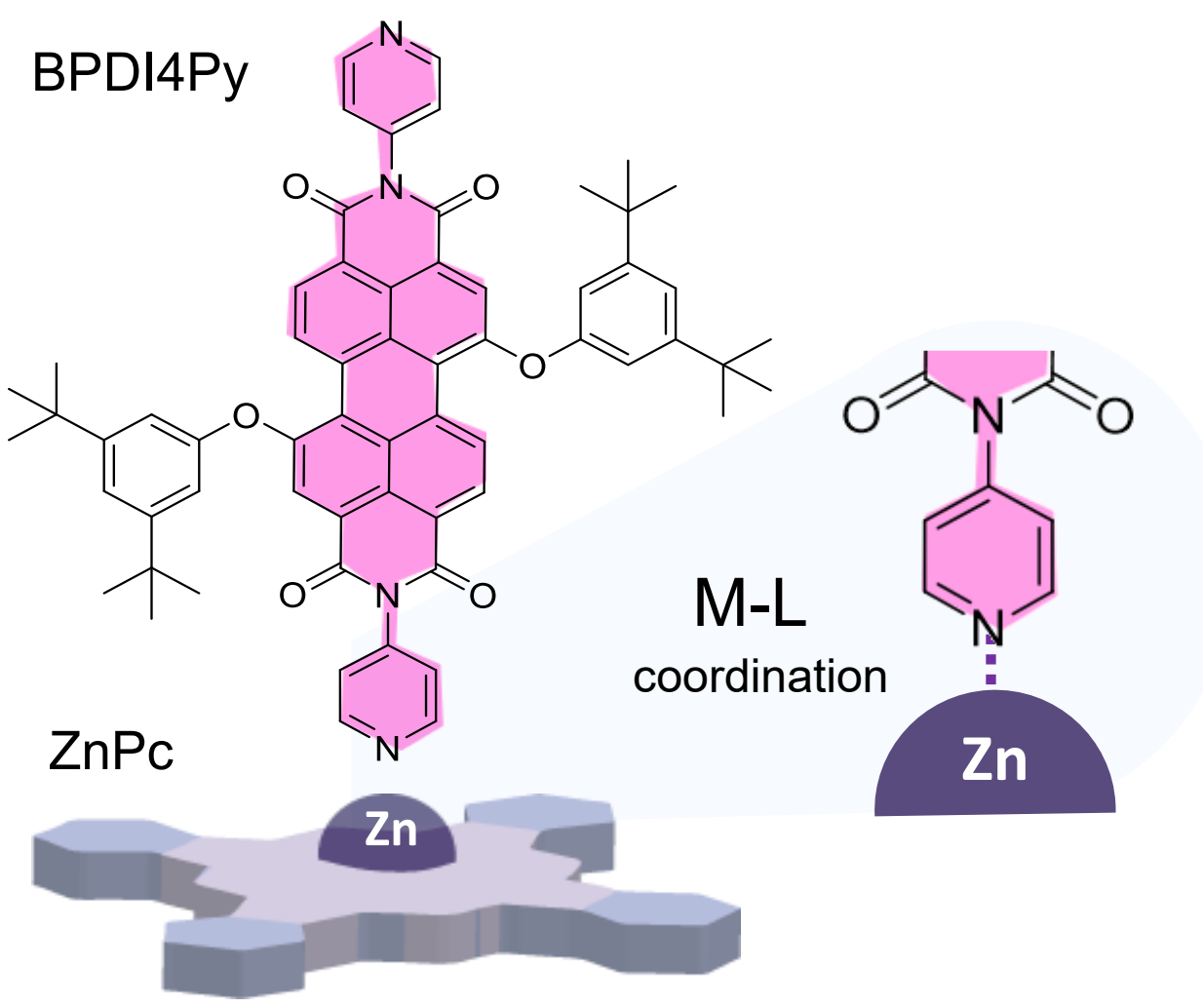

**Figure 5:** Molecular structures of the BPDI4Py molecule and scheme of the metal-ligand (M-L) complex formed with ZnPc (see details in the cited reference) [50].

In order to support our interpretation, we thus compared the immobilization of BPDI4Py onto a self-assembled ZnPc:TSB35-C12 monolayer on CVD graphene without and with annealing (3 h at 80°C). The presence of BPDI4Py after rinsing was checked by optical absorption micro-spectroscopy as reported in Figure 6. The wide incidence solid angle (0.8 N.A. see methods) allows being sensitive to the various molecular orientations. As expected, the red shift of the ZnPc Q-band is clearly visible after annealing. A new peak, not present in Figure 1, appears at 526 nm which can be attributed to the BPDI4Py $\pi$-$\pi^*$ optical transition [50]. This peak is clearly visible in the case of the annealed ZnPc:TSB35-C12 monolayer whereas it is hardly discernible in the sample deprived from prior annealing. This points to an easier formation of metal-ligand complex with ZnPc and pyridyl group after annealing, and supports an improved accessibility of the Zn atom. This could also explain the observation in the previous work [50] where in some cases the supramolecular dyad had to be pre-assembled in solution prior to deposition when the monolayers were not annealed.

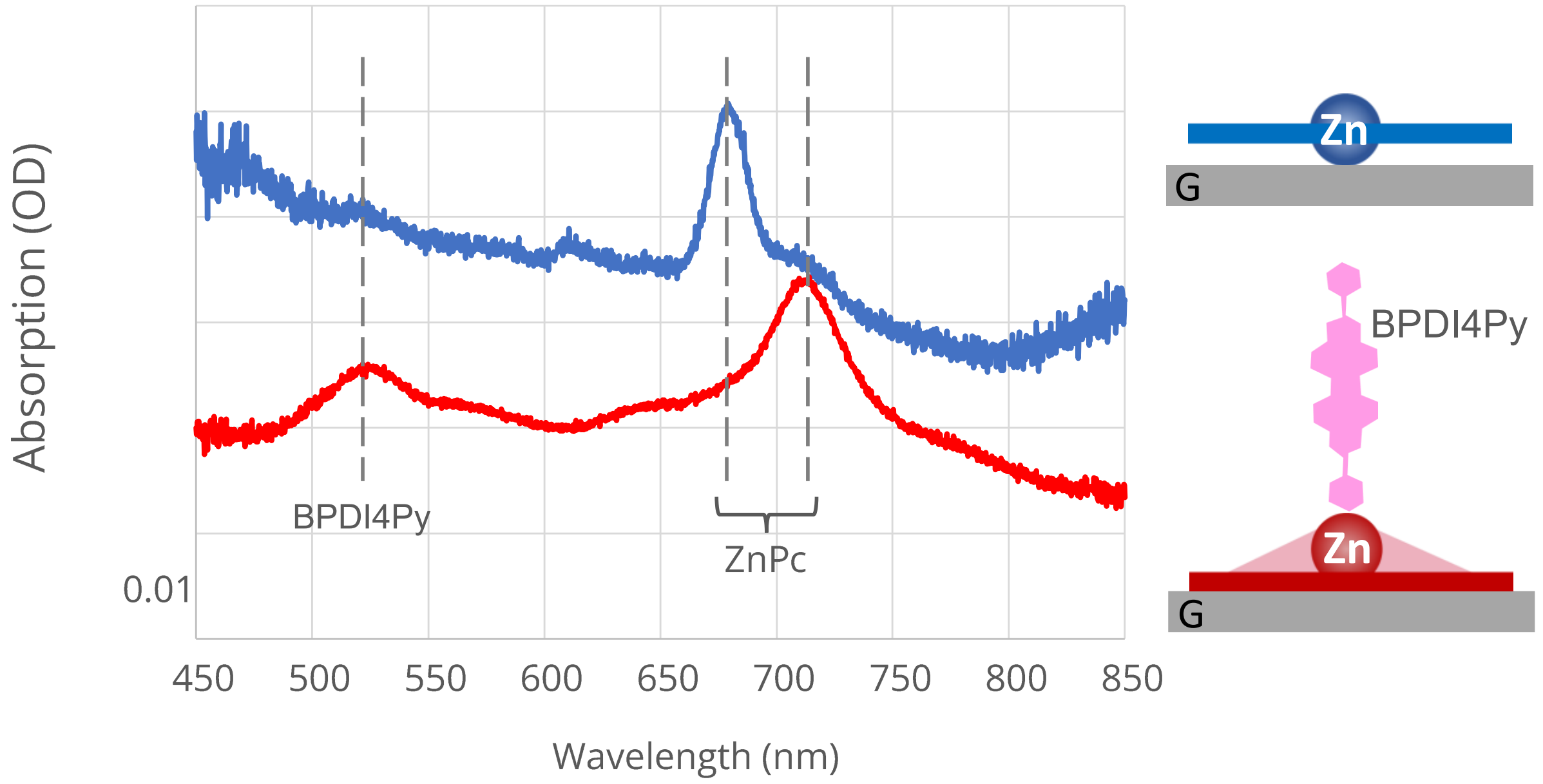

**Figure 6:** Left: Micro-absorption spectra of an annealed (3 h at 80°C, red line) or not-annealed (blue line) ZnPc:TSB35-C12 self-assembled monolayer on CVD graphene after its exposition to a solution of BPDI4Py. The peak at 526 nm is assigned to BPDI4Py. The curves have been vertically shifted for clarity. Right: scheme of the proposed interpretation.

# Conclusion

In summary, we have studied the effect of annealing on ZnPc molecules adsorbed on HOPG or graphene, isolated inside the single-molecule pores of a TSB35-C12 self-assembled matrix. We have shown that a moderate annealing for 3 h at 80°C produces a transition between two distinct adsorbed structures. This transition is studied through a combination of spectral and structural characterization techniques (absorption, Raman micro-spectroscopy, STM, absorption anisotropy) which reveal marked changes upon annealing. We tentatively interpret those changes as a transition of the adsorbed ZnPc molecules, individualized inside TSB35-C12 pores, from a planar structure dominated by Zn-substrate interactions towards a shuttlecock structure, with the Zn atom pointing outward, dominated by the interactions of $\pi$-conjugated PC moiety with the substrate.

This result is important in the context of bottom-up fabrication through molecular self-assembly. Actually, the adsorbed ZnPc emerges as a very valuable 3D-enabled tecton through ligand-metal bonding. [50] We have shown here that this process can be favored by a prior annealing of the ZnPc monolayer. The textbook case of ZnPc illustrates the possible role of annealing on bottom-up building of self-assembled molecular architectures. These results could also open important outlooks in the context of future organic electronics, since numerous electronic processes take place at the interface between substrate and molecule, and a proper control of its molecular-scale structure may be determinant [13].

# Experimental

## Sample preparation

The 2D guest-host ZnPc:TSB35-C12 samples were formed following the method detailed previously (monolayer case) [40,50]. In summary, the substrates were either freshly cleaved HOPG (SPI supplies, grade 2) or monolayer CVD graphene transferred from its growth Cu substrate with PMMA Coating (Graphenea) on a 170 µm thick transparent microscope glass cover plate.

The nanoporous 2D host network was formed by self-assembly of 1,3,5-tristyrylbenzene substituted in positions 3 and 5 by alkoxy peripheral chains presenting 12 carbon atoms (TSB35-C12) [39] and the guest molecule was Zinc Phthalocyanine (ZnPc, Sigma-Aldrich). The host-guest monolayers were successively drop-casted from toluene solutions with ~ 2:1 relative molar concentrations, respectively. The coordination with BPDI4Py was conducted in situ, on a fresh guest-host ZnPc:TSB35-C12 sample after complete solvent evaporation. A ~8µl droplet of an excess (typically $10^{-5}$M) solution of BPDI4Py in toluene was applied on the sample for ~5 min under a watch glass cover. The sample was then rinsed by dipping in toluene for ~1 min to remove unbonded BPDI4Py. The resulting sample was not re-annealed before optical measurements.

## Annealing

The annealing temperature was limited to 80°C, which appeared compatible with the stability of the TSB35-C12 nanoporous phase [40]. The annealing took place under ambient air conditions, the sample being simply covered by a watch glass.

## Optical micro-spectroscopy

Optical micro-spectroscopy was performed using an inverted optical microscope (IX71, Olympus) equipped with an objective lens (Nikon, 60×, 0.8 N.A.). The detection chain was made of an imaging spectrometer (Kymera 193i, Andor) equipped with a blazed grating (SR2-GRT-0150-0500, Andor) and a CMOS camera (Zyla 5.5, Andor). An automated slit was positioned at the entrance of the spectrometer and tuned to a width of 10 µm. Hyperspectral 1D-images were obtained with a spectrum acquired every 0.1 µm along a 230 µm long segment on the sample.

**Micro-absorption** was measured in transmission configuration with a white thermal lamp (12V 100W HAL-L, Olympus) as input illumination. The reference was acquired on a clean glass cover plate. The transmitted signal was collected through the objective lens and sent to the detection chain to construct a 1D hyperspectral image reported either in dark-background-corrected absorption ($100\times(1 - T / T_{reference})$, %) or in absorption $OD=-\log_{10}(T / T_{reference})$.

**Raman scattering** was measured together with residual fluorescence using a 633 nm excitation (60 µW $cm^{-2}$) from a stabilized laser (Cobolt 08-NLD series, Hubner Photonics). This excitation light was line-filtered (LL01-633, Semrock) and focused on the sample through the objective lens (Nikon, 60×, 0.8 N.A.). The retro-emitted light (Raman scattering and residual fluorescence) was collected through the microscope objective, filtered with appropriate long-pass filters (LPD02-633-RU + LP03-633-RU, Semrock) and sent to the above spectrometer. An instrumental background (*i.e.* acquired with a clean-glass sample) is systematically subtracted.

## Scanning Tunneling Microscopy

The STM images of the dry samples were recorded under ambient conditions (at air and room temperature $T \approx 300$ K) with a homemade digital system. The tip was mechanically cut in a Pt/Ir 250-µm wire (90/10, Goodfellow). The scanning piezoelectric ceramic was calibrated in the xy directions with the help of atomically resolved pictures obtained on HOPG. In order to ensure a correct statistical representation of our measurements concerning the structural organization of the monolayers, several images were systematically recorded at different locations of the sample. The images were acquired in quasi constant current mode (*i.e.* variable height mode). The setpoint tunneling current was ~3 pA and the sample bias voltage was -0.9 V. . The images are not corrected for the thermal drift.

### Incidence-angle resolved absorption.

Absorption spectra at various incidence angle were recorded on a QE-Pro Ocean Optics spectrophotometer. White light (MWWHL4, ThorLabs) was expanded and collimated using an achromatic doublet (AC254-100-A-ML, ThorLabs). Then, a diaphragm was used to obtain a beam size of 1 mm diameter on the sample. The polarization was controlled using a linear polarizer (LPVISC100, ThorLabs) to obtain *p* (*i.e.* Transverse-Magnetic) polarization. The incidence angle was tuned by rotating the sample holder from 0 to 65° (5° increments) and the spectrum was recorded at each angle, normalized for each angle with the transmission of a reference neat cover plate.

## Funding

The authors acknowledge the support of the French Agence Nationale de la Recherche (ANR), under grant ANR21-CE06-0041 (project LESOMMETA).